\documentclass[aps,pra,showpacs,superscriptaddress]{revtex4}%
\usepackage{amsmath,amssymb}

\begin{document}

\title{Nonlocality  without inequalities for almost all entangled
   states of any quantum system.}\thanks{Work supported in part by Istituto Nazionale di
  Fisica Nucleare, Sezione di Trieste, Italy.}

\author{GianCarlo \surname{Ghirardi}}\email{ghirardi@ts.infn.it}%
\affiliation{Department of Theoretical Physics of the University of
  Trieste, Italy}%
\affiliation{Istituto Nazionale di Fisica Nucleare, Sezione di Trieste,
  Italy}%
\affiliation{International Centre for Theoretical Physics ``Abdus Salam,''
  Trieste, Italy}%

\author{Luca \surname{Marinatto}}\email{marinatto@ts.infn.it}%
\affiliation{Department of Theoretical Physics of the University of
  Trieste, Italy}%
\affiliation{Istituto Nazionale di Fisica Nucleare, Sezione di Trieste,
  Italy}%

\date{\today}

\begin{abstract}
It is shown that it is possible to rule out all local and stochastic hidden variable
 models accounting for the quantum mechanical predictions implied by almost any
 entangled quantum state vector of any number of particles whose Hilbert spaces
 have arbitrary dimensions, without resorting to Bell-type inequalities.
The present proof makes use of the mathematically precise notion of Bell locality
 and it involves only simple set theoretic arguments.
\end{abstract}


\maketitle
%--------------------------------------------------------------------------

In a seminal paper~\cite{hardy}, L. Hardy has shown that for all
  entangled, but not maximally entangled, states of two spin-$1/2$ particles,
  one can prove nonlocality without resorting to inequalities.
More precisely he proved that ``local realism" unavoidably conflicts with the quantum predictions  by resorting
 to typical EPR-counterfactual arguments~\cite{epr}. Hardy's paper can be
 considered the best example of the attempts of proving nonlocality without using inequalities and, as such, it
 has been defined ``the best version of Bell's theorem''~\cite{merm}. Such an approach was initiated by the
 celebrated GHZ argument~\cite{ghz} with reference to a precise entangled state of three spin-$1/2$ particles.
Subsequently S. Goldstein~\cite{gold} simplified the proof of Ref.~\cite{hardy} and extended it to the case of
 bipartite systems whose constituents belong to Hilbert spaces of arbitrary dimensions.
Later on, new refinements~\cite{hardy1} of the original argument by Hardy,
 trying to maximize the probability of certain quantum mechanical outcomes which cannot be accounted for
 by any local deterministic hidden variable model, have been presented.

The aim of this paper is to reformulate and generalize Hardy and Goldstein's arguments
 by exhibiting, without resorting to inequalities, a contradiction between the existence
 of a local stochastic hidden variable model and the quantum mechanical predictions for almost any
  non-completely factorizable state vector of a system composed of an arbitrary
  number of particles whose Hilbert spaces have arbitrary dimensionality.
Our formal approach, inspired by the works
  of Refs.~\cite{hardy,gold},
  is particularly simple and straightforward since it involves only
elementary set theoretic
  arguments avoiding the use of counterfactuals, and it is completely general.
For what concerns the experimental implications, our approach has the advantage of
  identifying the precise set of measurements which can yield, with a
precise probability,
  outcomes whose occurrence put into evidence the unavoidable ``non-locality"
  of every conceivable hidden variable completion of quantum mechanics.
Contrary to all the nonlocality tests based on a violation of some Bell's inequality
  which require different correlation experiments, the tests
  based on a Hardy-like argument require simply the occurrence of a
unique and particular
  joint event, which is unambiguously identified once
one knows the
  state he is supplied with.

%---------------------------------------------------------------

Let us start by making precise the
   formal framework we are dealing with and the locality request we will use.
The framework will be the one of the so-called stochastic hidden variable theories for
 correlation experiments of the EPR-Bell type which involve measurement processes performed on a
   $n$-partite quantum system in the state
   $\vert \psi(1,2,\dots,n) \rangle \in {\cal H}_{1}\otimes
   \dots\otimes {\cal H}_{n}$ (where the dimensionality of ${\cal H}_{i}$ is
   arbitrary).
   It consists of: (i) a set $\Lambda$ whose elements
$\lambda$ are called hidden variables; (ii) a normalized and positive probability
  distribution $\rho$ defined on $\Lambda$; (iii)  a set of
probability distributions
   $P_{\lambda}(A_{i}\!=\!a, B_{j}\!=\!b,\dots, Z_{k}\!=\!z)$
   for the outcomes of single and joint measurements of any conceivable
   set of observables $\left\{A_{i},B_{j},\dots,Z_{k} \right\}$
   where each index $\left\{ i,j,\dots,k\right\}$ refers to a single
particle or to
   a group of particles, such that:
\begin{equation}
\label{eq0.1} P_{\psi}(A_{i}=a,B_{j}=b,\dots,Z_{k}=z) = \int_{\Lambda}\,d\lambda\,
   \rho(\lambda) P_{\lambda}(A_{i}=a,B_{j}=b,\dots,Z_{k}=z).
\end{equation}
Here the quantities at the left hand side are the probability distributions
   which quantum mechanics attaches to the outcomes
   $\left\{ a,b, \dots, z\right\}$ of the
   considered measurements when the system is in the state $\vert \psi \rangle$.
   A deterministic hidden variable model is a particular instance of a
stochastic one
   where all probabilities $P_{\lambda}$ can take only the values $0$ or $1$.

The considered framework is extremely general and the parameters $\lambda$ may be
   completely or only partially accessible, both practically or in principle,
   to the experimenter.
%For example the $\lambda$s might reduce to the state vector
%   $\vert \psi \rangle$ of the system, as happens in ordinary quantum mechanics,
%   and in this case no average procedure over the distribution $\rho(\lambda)$
%   is necessary, or they might represent both the state vector
%   $\vert \psi \rangle$ and the positions of the
%   particles of the system, as happens in Bohmian mechanics.
%In this last case, while $\vert \psi\rangle$ is considered as accessible,
%   the positions are assumed to be in principle not accessible so that
% an average
%   over their distribution is needed in order to reproduce the quantum
%   predictions.
%In the most general case, the accessible knowledge about the actual
%   physical situation can reduce simply to the knowledge of $\rho(\lambda)$
%   itself and the physical predictions are basically obtained by an averaging
%   procedure.
%An example of such a situation is represented by Bell's model identifying
%   the states of a spin-$1/2$ particle by unit vectors in a tridimensional
%   Euclidean space~\cite{bell1}.

In the case in which the measurement processes take place at spacelike
   separated locations, the following condition demanding that
   all conceivable probability distributions of measurement processes
   satisfy the factorization property
\begin{equation}
\label{eq0.2}
   P_{\lambda}(A_{i}=a,B_{j}=b,\dots,Z_{k}=z) = P_{\lambda}(A_{i}=a)
   P_{\lambda}(B_{j}=b)\dots P_{\lambda}(Z_{k}=z)  \hspace{1cm}\forall
\lambda\in \Lambda\:,
\end{equation}
is a physically natural one which every hidden variable model is
   requested to satisfy.\\
This factorizability request is commonly known as Bell's locality condition~\cite{bell}. We remark that all
``nonlocality without inequalities" proofs aim at exhibiting
  a conflict between the quantum predictions for a specific entangled state
  and any local completion of quantum mechanics which goes beyond
quantum mechanics itself. In fact, in the particular case in which the most complete specification of the
  state of a physical system is represented by the knowledge of the state vector
  $\vert \psi \rangle$ alone, i.e., within ordinary quantum mechanics,
the failure
  of the locality condition of Eq.~(\ref{eq0.2}) can be established
directly by plugging
  into it appropriate quantum mechanical observables.
Indeed, it is a well-known fact that for any entangled state there exist
  joint probabilities which do not factorize and, consequently, that ordinary
  quantum mechanics is inherently a nonlocal theory.

As it has been noted by A. Fine~\cite{fine}
  local stochastic hidden variable models are  equivalent to local
  deterministic hidden variable models.
Nonetheless, in what follows we prefer working within the stochastic scenario. This choice, which differs from
the one usually made in the literature~\cite{hardy,ghz,gold,hardy1}, allows us to  express the locality request
as the factorizability condition on joint
   probabilities instead of resorting to Einstein locality~\cite{epr}.

%--------------------------------------------------------------------

To start with, we consider an $n$-partite system described by the state vector $\vert \psi(1,\dots,n)\rangle$
  belonging to the Hilbert space ${\cal H}_{1}\otimes \dots
\otimes{\cal H}_{n}$---where the dimension
  of ${\cal H}_{i},\:i=1,\dots ,n$ is greater than or equal to $2$.
Let us split arbitrarily the $n$ particles in two subsets, which we will label for convenience
  as $1$ and $2$ in what follows, and suppose they involve particles
laying into two space-like separated
  spatial regions. Finally, let us consider the Schmidt decomposition
of $\vert \psi\rangle$
  in terms of the orthonormal sets of states $\left\{ \vert
\alpha_{i}(1) \rangle \right\}$
  and $\left\{ \vert \beta_{i}(2) \rangle \right\}$, referring to the
first and to the second
  group of particles respectively:
\begin{equation}
\label{eq2} \vert \psi(1,2)\rangle = \sum_{i}p_{i}\vert \alpha_{i}(1)\rangle \otimes
   \vert \beta_{i}(2) \rangle\:,
\end{equation}
where the positive weights $p_{i}$ satisfy the normalization condition
   $\sum_{i}p_{i}^{2}=1$.
Suppose now that in Eq.(\ref{eq2}) at least two different weights appear which we assume for simplicity to
   be the first two, so that $p_{1}\neq p_{2}$.
Actually, this is the only hypothesis which is required by our proof. This means that the nonlocality without
inequalities proof we are going to present holds for those non-completely
  factorizable states whose Schmidt coefficients are not all
equal~\footnote{Accordingly, and curiously enough,
  a proof of the kind we are considering does not apply, for instance,
to the maximally
  entangled states, the ones which exhibit the maximum amount of
nonlocal correlations.}. In terms of the coefficients $p_{1}$ and $p_{2}$, we build up the following two
$2\times 2$ unitary matrices
   $U$ and $V$:
\begin{equation}
\label{eq3} U=\frac{1}{\sqrt{p_{1}+p_{2}}}
\begin{bmatrix}
\sqrt{p_{2}} & -i\sqrt{p_{1}} \\
-i \sqrt{p_{1}} & \sqrt{p_{2}}
\end{bmatrix}
V=\frac{1}{\sqrt{p_{1}^{2}+p_{2}^{2}-p_{1}p_{2}}}
\begin{bmatrix}
-i(p_{2}-p_{1}) & \sqrt{p_{1}p_{2}} \\
\sqrt{p_{1}p_{2}} & -i(p_{2}-p_{1})
\end{bmatrix}\:.
\end{equation}
We define now two orthonormal bases
   $\left\{ \vert x_{+}(1) \rangle, \vert x_{-}(1)\rangle \right\}$ and
   $\left\{ \vert y_{+}(1) \rangle, \vert y_{-}(1)\rangle \right\}$ of the
   two-dimensional linear manifold of the first group of particles
spanned by the vectors
   $\left\{ \vert \alpha_{1}(1)\rangle,\vert \alpha_{2}(1)\rangle\right\}$, and
   two bases $\left\{ \vert x_{+}(2) \rangle, \vert x_{-}(2)\rangle
\right\}$ and
   $\left\{ \vert y_{+}(2)\rangle, \vert y_{-}(2)\rangle \right\}$ of
   the two-dimensional linear manifold of the second group of
particles spanned by
   the vectors
   $\left\{ \vert \beta_{1}(2)\rangle,\vert \beta_{2}(2)\rangle\right\}$,
   according to:
\begin{equation}
\label{eq4}
\begin{bmatrix} \vert x_{+}(1) \rangle \\  \vert x_{-}(1) \rangle \end{bmatrix}
=U \begin{bmatrix} \vert \alpha_{1}(1) \rangle \\  \vert \alpha_{2}(1) \rangle
\end{bmatrix} \hspace{1cm}
\begin{bmatrix} \vert y_{+}(1) \rangle \\  \vert y_{-}(1) \rangle \end{bmatrix}
=VU \begin{bmatrix} \vert \alpha_{1}(1) \rangle \\  \vert \alpha_{2}(1)
   \rangle
\end{bmatrix} \hspace{0.1cm}
\end{equation}
\begin{equation}
\label{eq4.01}
\begin{bmatrix} \vert x_{+}(2) \rangle \\  \vert x_{-}(2) \rangle \end{bmatrix}
=U \begin{bmatrix} \vert \beta_{1}(2) \rangle \\  \vert \beta_{2}(2) \rangle
\end{bmatrix} \hspace{1cm}
\begin{bmatrix} \vert y_{+}(2) \rangle \\  \vert y_{-}(2) \rangle \end{bmatrix}
=VU \begin{bmatrix} \vert \beta_{1}(2) \rangle \\  \vert \beta_{2}(2) \rangle
\end{bmatrix} \:.
\end{equation}
 From Eqs.~(\ref{eq4}-\ref{eq4.01}), one immediately sees that the state
   $\vert \psi \rangle$ of Eq.~(\ref{eq2}) can be expressed
   in the following three equivalent forms:
\begin{eqnarray}
\label{eq5}
   \vert \psi(1,2)\rangle & = & i\sqrt{p_{1}p_{2}}\,[\,\vert x_{+}(1)\rangle
   \vert x_{-}(2) \rangle + \vert x_{-}(1)\rangle  \vert x_{+}(2) \rangle\,]
   +(p_{2}-p_{1}) \vert x_{-}(1)\rangle \vert x_{-}(2) \rangle
   + \sum_{i>2} p_{i}\vert \alpha_{i}(1)\rangle \vert \beta_{i}(2) \rangle
   \nonumber\\
   & = & i\sqrt{p_{1}^{2}+p_{2}^{2}-p_{1}p_{2}}\, \vert y_{-}(1)\rangle
   \vert x_{-}(2) \rangle +i \sqrt{p_{1}p_{2}}\,\vert x_{-}(1)\rangle \vert
   x_{+}(2) \rangle + \sum_{i>2} p_{i}\vert \alpha_{i}(1)\rangle \vert
   \beta_{i}(2) \rangle \nonumber\\
   & = & i\sqrt{p_{1}p_{2}}\,[\,\vert x_{+}(1)\rangle
   \vert x_{-}(2) \rangle + i\sqrt{p_{1}^{2}+p_{2}^{2}
-p_{1}p_{2}} \vert x_{-}(1)\rangle \vert
   y_{-}(2)  \rangle+ \sum_{i>2} p_{i}\vert \alpha_{i}(1)\rangle \vert
   \beta_{i}(2) \rangle\:.
\end{eqnarray}
Let us now denote by $X_{1},Y_{1},X_{2}$ and $Y_{2}$ the four
   observables, having as eigenstates associated to the
   eigenvalues $+1$ and $-1$ the previously defined couples of orthogonal
   vectors $ \left\{ \vert x_{+}(1) \rangle, \vert x_{-}(1)\rangle \right\}$,
$\left\{ \vert y_{+}(1) \rangle, \vert y_{-}(1)\rangle \right\}$ and $\left\{ \vert x_{+}(2) \rangle, \vert
x_{-}(2)\rangle \right\}$,
   $\left\{ \vert y_{+}(2) \rangle, \vert y_{-}(2)\rangle \right\}$
respectively,
   while they act as the null operator in the manifolds
   orthogonal to the bidimensional ones corresponding to the non-zero
   eigenvalues.
According to Eq.~(\ref{eq5}) and taking into account the orthogonality of the
   involved states,  the quantum joint probabilities concerning the
   set of observables $X_{1},Y_{1},X_{2}$ and $Y_{2}$ satisfy the
   following relations:
\begin{eqnarray}
\label{eq7.1}
P_{\psi}(X_{1}=+1, X_{2}=+1) &=& 0\\
\label{eq7.2}
P_{\psi}(Y_{1}=+1, X_{2}=-1) &=& 0 \\
\label{eq7.3}
P_{\psi}(X_{1}=-1, Y_{2}=+1) &=& 0 \\
\label{eq7.31}
P_{\psi}(Y_{1}=+1, X_{2}=0) &=& 0\\
\label{eq7.32}
P_{\psi}(X_{1}=0, Y_{2}=+1) &=& 0 \\
\label{eq7.4} P_{\psi}(Y_{1}=+1, Y_{2}=+1) &\neq&  0 \:.
\end{eqnarray}
Suppose now that a local stochastic hidden variable model reproducing, according to Eq.~(\ref{eq0.1}),
   the quantum predictions for the state $\vert \psi \rangle$, exists.
As a consequence, considering for example Eq.~(\ref{eq7.1}), we must have:
\begin{eqnarray}
\label{eq7.5} P_{\psi}(X_{1}=+1, X_{2}=+1) &  = & \int_{\Lambda}d\lambda \rho(\lambda)
P_{\lambda}(X_{1}=+1, X_{2}=+1) \nonumber \\
&= & \int_{\Lambda} d\lambda \rho(\lambda) P_{\lambda}(X_{1}=+1)P_{\lambda}(X_{2}=+1)= 0\:,
\end{eqnarray}
where the second equality follows from the request that our model
   satisfies Bell's locality condition of Eq.~(\ref{eq0.2}).
Since $\rho(\lambda)$ is a positive distribution, the last integral of
   Eq.~(\ref{eq7.5}) vanishes if and only if the product
   $P_{\lambda}(X_{1}=+1)P_{\lambda}(X_{2}=+1)$ vanishes almost
   everywhere~\footnote{By the expression ``almost everywhere'' it is meant
   that the argument of the integral may be different from zero at most within a
   non-empty set $\Gamma$ such that $\int_{\Gamma}d\lambda \rho(\lambda)=0$.}
   within $\Lambda$.
Analogous reasonings can be applied to Eqs~(\ref{eq7.2}-\ref{eq7.4}),
   leading to:
\begin{eqnarray}
\label{eq8.1}
P_{\lambda}(X_{1}=+1)P_{\lambda}(X_{2}=+1) &=& 0 \\
\label{eq8.2}
P_{\lambda}(Y_{1}=+1)P_{\lambda}(X_{2}=-1) &=& 0 \\
\label{eq8.3}
P_{\lambda}(X_{1}=-1)P_{\lambda}(Y_{2}=+1)  &=& 0 \\
\label{eq8.31}
P_{\lambda}(Y_{1}=+1)P_{\lambda}(X_{2}=0) &=& 0 \\
\label{eq8.32}
P_{\lambda}(X_{1}=0)P_{\lambda}(Y_{2}=+1) &=& 0\\
\label{eq8.4} P_{\lambda}(Y_{1}=+1)P_{\lambda}(Y_{2}=+1) &\neq & 0  \:,
\end{eqnarray}
where the first five equations must hold almost everywhere within $\Lambda$,
   while the sixth equation has to be satisfied in a subset of
   $\Lambda$ whose measure according to the distribution $\rho(\lambda)$ is
   non-zero.
It is worth noticing that, in the special case where the parameters
   $\lambda$ coincide with the state vector $\vert \psi \rangle$
   alone, as happens in ordinary quantum mechanics, the previous equations
   are immediately seen to be violated.
In fact, for the state $\vert \psi \rangle$ of Eq.~(\ref{eq5})
   both $P_{\lambda=\psi}(X_{1}=+1)$ and $P_{\lambda=\psi}(X_{2}=+1)$ are
   different from zero, thus contradicting Eq.~(\ref{eq8.1}).

We prove now, with simple set manipulations, that no conceivable local stochastic
  hidden variable model exists which is compatible with the
  previous equations.
To this end, let us first of all notice that by summing Eq.~(\ref{eq8.2})
  and~(\ref{eq8.31}) and taking into account that $P_{\lambda}(X_{2}=-1)+
  P_{\lambda}(X_{2}=+0)+P_{\lambda}(X_{2}=+1)=1$ we have:
\begin{equation}
\label{eq8.41} P_{\lambda}(Y_{1}=+1)[1-P_{\lambda}(X_{2}=+1)]=0\:.
\end{equation}
Similarly, summing Eqs.~(\ref{eq8.3}) and~(\ref{eq8.32}) we have:
\begin{equation}
\label{eq8.42} [1-P_{\lambda}(X_{1}=+1)]P_{\lambda}(Y_{2}=+1)=0\:.
\end{equation}
Now we define the following subsets $A$, $B$ and $C$ of $\Lambda$ as:
\begin{eqnarray}
   A &=& \left\{ \lambda\in \Lambda  \vert P_{\lambda}(X_{1}=+1)=0  \right\}, \\
  B & =&\left\{ \lambda\in \Lambda  \vert P_{\lambda}(X_{2}=+1)=0\right\},\\
  C &=& \Lambda-(A\cup B).
\end{eqnarray}
Since $\Lambda-(A\cup B)= (\Lambda-A)\cap (\Lambda-B)$, we have that,
   for all $\lambda$ belonging to $C$, $P_{\lambda}(X_{1}=+1)
   P_{\lambda}(X_{2}=+1)\neq 0$.
If the set $C$ would have a non-zero measure according to the
   distribution $\rho$, i.e., if $\int_{C}d\lambda \rho(\lambda)\neq 0$,
   one would violate Eq.~(\ref{eq8.1}) and, consequently, Eq.~(\ref{eq7.1}).
Therefore, in order to fulfill Eq.~(\ref{eq8.1}), the set $A\cup B$
   must coincide with $\Lambda$ exception made for a set of zero measure.
We  can then argue as follows. If $\lambda$ belongs to $A$ then, by definition, $P_{\lambda}(X_{1}=+1)=0$,
   so that, according to Eq.~(\ref{eq8.42}) we get $P_{\lambda}(Y_{2}=+1)=0$.
Equivalently, if $\lambda$ belongs to $B$ then $P_{\lambda}(X_{2}=+1)=0$ and,
   according to Eq.~(\ref{eq8.41}), $P_{\lambda}(Y_{1}=+1)=0$.
Therefore, for any $\lambda \in A\cup B$
   either $P_{\lambda}(Y_{1}=+1)= 0$ or $P_{\lambda}(Y_{2}=+1)= 0$,
   contradicting Eq.~(\ref{eq8.4}).
We have thus proven that it does not exist a subset of $\Lambda$ of non-zero
   measure such that Eq.~(\ref{eq8.4}) can be satisfied.

This simple argument shows that it is not possible to exhibit any conceivable
 stochastic hidden variable model, satisfying Bell's locality condition, which
 can account for the quantum mechanical predictions of almost any $n$-partite
 quantum entangled state $\vert \psi \rangle \in {\cal H}_{1}\otimes \dots \otimes {\cal H}_{n}$,
 whenever there exists a splitting of the $n$ particles whose Schmidt decomposition
 contains at least two different weights.

The experimental test of nonlocality for the considered set of entangled states consists simply in testing the
occurrence of the joint
 measurement outcomes which Eq.~(\ref{eq7.4}) shows to be possible. Given the state of Eq.~(\ref{eq5}), it
 is straightforward to show that this event happens with a probability:
\begin{equation}
 \label{eq8.5}
 P_{\psi}(Y_{1}=+1, Y_{2}=+1) = \frac{p_{1}^{2}p_{2}^{2}(p_{1}-p_{2})^2}{(p_{1}^{2}+p_{2}^{2}-p_{1}p_{2})^{2}}
\end{equation}
which does not vanish whenever $p_{1}, p_{2} \neq 0$ and $p_{1} \neq p_{2} $.

%-----------------------------------------------------------------------

The joint probability distribution of Eq. (\ref{eq8.5}) involves measurements of the observables $Y_{1}$ and
$Y_{2}$, which are generally multipartite. As well known, from a strict practical point of view, the measurement
of such observables might turn out to be not feasible and it would be in any case quite difficult to perform. In
order to overcome this problem, we are going to exhibit now
  a variant of the previous proof which makes reference to repeated
application of the Schmidt decomposition of the $n$-partite state in
  order to identify a non-zero joint probability distribution which
conflicts with Bell's locality condition and which  involves only single-particle
  observables.

To accomplish this task, we  start by considering a tripartite system and then we
    show how to generalize the procedure to any number of particles.
Consider the Schmidt decomposition of a tripartite state $\vert \psi(1,2,3)
   \rangle$ in terms of a set of bipartite orthonormal states of the first and
   the second particle $\left\{ \vert \phi_{k}(1,2)\rangle \right\}$,
   and of a set of orthonormal states of the third particle $\left\{ \vert
   \tau_{k}(3) \rangle \right\}$:
\begin{equation}
\label{eq10} \vert \psi(1,2,3)\rangle =\sum_{k} q_{k} \vert \phi_{k}(1,2)\rangle
   \otimes \vert \tau_{k}(3)\rangle.
\end{equation}
where $q_{k}\geq 0$ and $\sum_{k}q_{k}^{2}=1$. Suppose now that, within the orthonormal set
   $\left\{\vert \phi_{k}(1,2)\rangle \right\}$
   there exists a state, whose associated $q_{k}$ is different from zero,
   let us say $\vert \phi_{1}(1,2)\rangle$,
   such that at least two different weights appear in its Schmidt
decomposition~\footnote{Actually the
   tripartite states for which our proof holds, are the ones whose
   Schmidt decomposition, performed by grouping the particles in every
   possible manner, contains at least one bipartite entangled
   state for an arbitrary pair of particles which satisfies the above request.}.
By resorting to the Schmidt decomposition of such a state and    performing the unitary transformations $U$ and
$V$ as defined in
   Eqs.~(\ref{eq4}-\ref{eq4.01}), the tripartite state $\vert
\psi(1,2,3) \rangle$
   exhibits the following form:\emph{}
\begin{eqnarray}
\label{eq11} \vert \psi(1,2,3)\rangle & = & q_{1}\Big( i\sqrt{p_{1}p_{2}}\,[\,
   \vert x_{+}(1)\rangle
   \vert x_{-}(2) \rangle + \vert x_{-}(1)\rangle  \vert x_{+}(2) \rangle\,]
   +(p_{2}-p_{1}) \vert x_{-}(1)\rangle \vert x_{-}(2) \rangle
   \Big) \otimes \vert \tau_{1}(3) \rangle \nonumber \\
& &  + \,q_{1}\Big( \sum_{i>2} p_{i}\vert \alpha_{i}(1)\rangle \vert
   \beta_{i}(2) \rangle \Big) \otimes \vert \tau_{1}(3) \rangle
    + \sum_{k>1} q_{k} \vert \phi_{k}(1,2) \rangle \otimes \vert
   \tau_{k}(3)\rangle\:,
\end{eqnarray}
which can be also rewritten in two other equivalent forms, in analogy
   with Eq.~(\ref{eq5}).

Denoting by $T_{3}$ the single-particle observable of the
   Hilbert space of the third particle having the vectors
   $\left\{ \vert \tau_{k}(3) \rangle \right\}$ as its eigenstates associated
   to a set of eigenvalues  $\left\{ t_{k} \right\}$ which we may choose
   so that $t_{1}$ is a non-degenerate eigenvalue, the following quantum
   probability distributions hold for the state of Eq.~(\ref{eq11}):
\begin{eqnarray}
\label{eq12.1}
P_{\psi}(X_{1}=+1, X_{2}=+1, T_{3}=t_{1}) &=& 0\\
\label{eq12.2}
P_{\psi}(Y_{1}=+1, X_{2}=-1, T_{3}=t_{1}) &=& 0 \\
\label{eq12.3}
P_{\psi}(X_{1}=-1, Y_{2}=+1, T_{3}=t_{1}) &=& 0 \\
\label{eq12.31}
P_{\psi}(Y_{1}=+1, X_{2}=0, T_{3}=t_{1}) &=& 0 \\
\label{eq12.32}
P_{\psi}(X_{1}=0, Y_{2}=+1, T_{3}=t_{1}) &=& 0 \\
\label{eq12.4} P_{\psi}(Y_{1}=+1, Y_{2}=+1, T_{3}=t_{1}) &\neq&  0 \:.
\end{eqnarray}
Once again the existence of a local and stochastic hidden variable model reproducing the quantum mechanical
probability distributions
   of Eqs.~(\ref{eq12.1}-\ref{eq12.4}), implies the following relations:
\begin{eqnarray}
\label{eq13.1}
P_{\lambda}(X_{1}=+1)P_{\lambda}(X_{2}=+1)P_{\lambda}(T_{3}=t_{1}) &=& 0 \\
\label{eq13.2}
P_{\lambda}(Y_{1}=+1)P_{\lambda}(X_{2}=-1)P_{\lambda}(T_{3}=t_{1}) &=& 0 \\
\label{eq13.3}
P_{\lambda}(X_{1}=-1)P_{\lambda}(Y_{2}=+1)P_{\lambda}(T_{3}=t_{1}) &=& 0 \\
\label{eq13.31}
P_{\lambda}(Y_{1}=+1)P_{\lambda}(X_{2}=0)P_{\lambda}(T_{3}=t_{1}) &=& 0 \\
\label{eq13.32}
P_{\lambda}(X_{1}=0)P_{\lambda}(Y_{2}=+1)P_{\lambda}(T_{3}=t_{1}) &=& 0 \\
\label{eq13.4}
  P_{\lambda}(Y_{1}=+1)P_{\lambda}(Y_{2}=+1)P_{\lambda}(T_{3}=t_{1})&\neq& 0\:.
\end{eqnarray}
As before, the first five equations must be satisfied almost everywhere
   within $\Lambda$, while the last one must be satisfied in a set of non-zero
   measure with respect to the distribution $\rho(\lambda)$.
Let us now show that a local stochastic hidden variable model
   satisfying Eqs.~(\ref{eq13.1}-\ref{eq13.4}) cannot exist.
The procedure we are going to follow  consists in splitting the set of hidden variables
   $\Lambda$ into two complementary and disjoint subsets $\Omega_{1}$ and
   $\Omega_{2}$, referring to the possible values of the probability
distribution
   for the outcome $t_{1}$ of the observable $T_{3}$.
They are defined as $\Omega_{1} =\left\{ \lambda\in \Lambda \vert
   P_{\lambda} (T_{3}=t_{1})=0\right\}$ and $\Omega_{2} =\left\{ \lambda\in
   \Lambda \vert  P_{\lambda} (T_{3}=t_{1})\neq 0\right\}$.
Given any value of the hidden variable $\lambda$, two possible cases can
   occur: either $\lambda \in \Omega_{1}$ or $\lambda \in \Omega_{2}$.
If $\lambda \in \Omega_{1}$, the left hand side of Eq.~(\ref{eq13.4}) vanishes
   and the equation cannot be satisfied.
If $\lambda$ belongs to $\Omega_{2}$ where $P_{\lambda}(T_{3}=t_{1})\neq 0$,
   the equations from~(\ref{eq13.1}) to~(\ref{eq13.4}) reduce to
   Eqs~(\ref{eq8.1}-\ref{eq8.4}) respectively.
This being the case, we can apply the previous arguments  to conclude that no local stochastic hidden variable
model   exists which can reproduce the quantum probabilities for all tripartite
   states, whose Schmidt decomposition involves at least one bipartite
   state having at least two different weights in its decomposition.
The argument can be generalized in
   a straightforward way to any number of particles.
Note that the $n$-particle entangled states for which our proof holds are
   those for which, by considering all conceivable Schmidt decompositions
   in terms of bipartite states and $(n-2)$ single-particle states,
   at least one entangled bipartite state involving at least two
different weights in its Schmidt decomposition
   appears.
Such a set contains almost all entangled states of $n$-particles. Once again, our approach uniquely determines
the set of outcomes of
  joint  measurements whose occurrence constitutes the
  experimental proof of the nonlocality of every conceivable hidden
  variable completion of quantum mechanics.
However, contrary to the first method we have  presented, it could be easily implemented in practice since it
involves only simple single-particle
  measurements.

%--------------------------------------------------------------------

\end{document}